\documentclass[10pt]{amsart}
\usepackage{amsmath}
\usepackage{amssymb}
\usepackage{yfonts}

\usepackage{tikz,pgfplots}

\usetikzlibrary{calc, patterns,arrows, shapes.geometric}

\begin{document}

\title{Notes on the flexible manipulator}
\author{Marcelo Epstein}

\address{Department of Mechanical and Manufacturing Engineering, University of Calgary,
Canada}
\email{mepstein@ucalgary.ca}


\date{}
\maketitle
\begin{abstract}
The existence of solutions to the boundary tracking of the displacement at one end of a linear Timoshenko beam is discussed on the basis of the Cauchy problem with time and space interchanged.
\end{abstract}

\section{Introduction}
Holding a fishing rod at one end ($x=0$) and starting from the rod at rest, is it possible to achieve any desired displacement at the other ($x=L$) by prescribing the motion of the held end? This problem has a long history and is of practical relevance for the control of flexible robotic manipulators, a field of research that has attained a high degree of mathematical sophistication. The modest objective of this note, on the other hand, is to shed some light on the feasibility of solutions of the problem from the point of view of the elementary theory of partial differential equations. The analysis is conducted within the realm of small deflections of a Timoshenko beam. Two separate situations are envisioned. In the first, that we call the 2-2 problem, to achieve any desired displacement and rotation functions at the free end it is required to prescribe both the displacement and the rotation of the held end. Since initially the rod is at rest, the motion at the free end will start with a delay dictated by the slower speed of wave propagation in the rod. The existence of a solution to this problem is argued on the basis of the existence of a solution to the Cauchy problem when the roles of time and space are interchanged. The second situation, called the 1-1 problem, consists of attaining a desired displacement evolution at the free end while keeping the displacement of the held end fixed and prescribing only its rotation. The solution in this case is based upon the solution of a recursive functional equation.

\section{Beam equations}
\label{sec:beamequations}
 Under fairly general assumptions, namely,
 \begin{enumerate}
 \item cross sections remain plane, though not necessarily perpendicular to the deformed axis of the beam,
 \item both translational and rotational inertia terms are included
 \item transverse deflections $w$ are very small compared with the length of the beam
 \item the material abides by Hooke's law
 \end{enumerate}
 Timoshenko\footnote{See, e.g., Timoshenko S P (1937), {\it Vibration Problems in Engineering}, Van Nostrand, p 337.} derived the dynamic equations of an ideal elastic beam, which bears his name. For the free motion (no external forces) of a beam of constant cross section and uniform material properties, Timoshenko's equations can be expressed in terms of the transverse deflection $w$ and the cross-section rotation $\phi$ as
 \begin{equation} \label{beam-1}
 Aw_{tt}=w_{xx}-\phi_x,
 \end{equation}
 and
 \begin{equation}\label{beam0}
 B \phi_{tt}=\phi_{xx}+C(w_x-\phi),
 \end{equation}
 where $A,B,C$ are positive constants and where subscripts indicate partial derivatives with respect to the natural body-time coordinates $x,t$.

It is possible to eliminate $\phi$ between the two equations so as to obtain a single fourth-order PDE for the transverse displacement. The result is
\begin{equation} \label{beam1}
w_{xxxx}-(A+B)w_{xxtt}+AB w_{tttt}+Dw_{tt}=0,
\end{equation}
where $D=CA$. The constants $A$, $B$ and $D$ represent, respectively, relative measures of the shear compliance, rotational inertia and translational inertia. In the limiting case when the stiffness compliance and the rotational inertial approach zero values, we recover the classical Bernoulli beam equation. An interesting intermediate case is that for which only the shear compliance vanishes.

It is also possible to express the equations of motion as a symmetric system of four first-order PDEs. The original formulation in terms of two second-order equations, however, is most suitable for the imposition of physically meaningful boundary conditions.

The characteristic polynomial\footnote{See, e.g., Courant R and Hilbert D (1962), {\it Methods of Mathematical Physics}, Interscience, Vol. II p 175.} associated with this differential equation is
\begin{equation} \label{beam2}
Q(\rho)=\rho^4-(A+B)\rho^2+AB.
\end{equation}
The roots of this polynomial are
\begin{equation} \label{beam3}
\rho_1=-\rho_2=\sqrt{A}
\end{equation}
and
\begin{equation} \label{beam4}
\rho_3=-\rho_4=\sqrt{B}.
\end{equation}
Discarding the unlikely case in which $A=B$,\footnote{Generally, $A<B$.} Equation (\ref{beam1}) is a totally hyperbolic PDE, with $v_i=1/\rho_i\;(i=1,...,4)$ as the four distinct characteristic speeds. Physically, $v_1$ and $v_2$ correspond, respectively, to the forward and backward bending waves, while $v_3$ and $v_4$ pertain to the shear waves. Notice that in the case of the Bernoulli beam all roots $\rho_i$ of the characteristic polynomial vanish, which indicates an infinite speed of propagation of all signals. In the intermediate case of a Bernoulli beam endowed with rotational inertia ($B>A=0$), bending waves propagate at a finite speed. The Bernoulli beam equation stands with respect to the Timoshenko beam in a relation somewhat analogous to the relation between the heat equation and the wave equation. The fact that all characteristic roots vanish implies not merely the loss of hyperbolicity but also that, the characteristic line being purely spatial, the initial value problem specifies initial data on a characteristic line, just as in the case of the classical heat equation. Although considerations of well-posedness can be derived for these non-hyperbolic characteristic-initial problems, the theory for totally hyperbolic problems is better established. For this reason, we will limit our considerations to the Timoshenko beam and adduce physical arguments to claim that the results obtained for a flexible manipulator based on the Timoshenko beam are also applicable to the Bernoulli beam.

\section{Well-posedness of initial and boundary-value problems}
\label{sec:wellposedness}

\begin{figure}
\begin{center}

\begin{tikzpicture}[scale = 0.5]
\draw[-stealth'] (-2,0) to  (10,0);
\draw[-stealth'] (0,-2) to  (0,10);
\node[right] at (10,0) {$x$};
\node[above] at (0,10) {$t$};
\draw[ultra thick] (-2,1) to [out=30,in=190] (10,5);
\node[right] at (10,5) {$\gamma$};
\node[above] at (4,7.5) {$P$};
\draw[thick] (4,7.5) to (1,2.5);
\draw[thick] (4,7.5) to (6.1,4);
\draw[thick] (4,7.5) to (2.5,3);
\draw[thick] (4,7.5) to (5.2,3.9);
\node[below] at (1,2.5) {$Q$};
\node[below] at (6.1,4){$R$};
\end{tikzpicture}
\end{center}
\caption{Domain of dependence $QR$ of point $P$}
\label{fig:domain}
\end{figure}
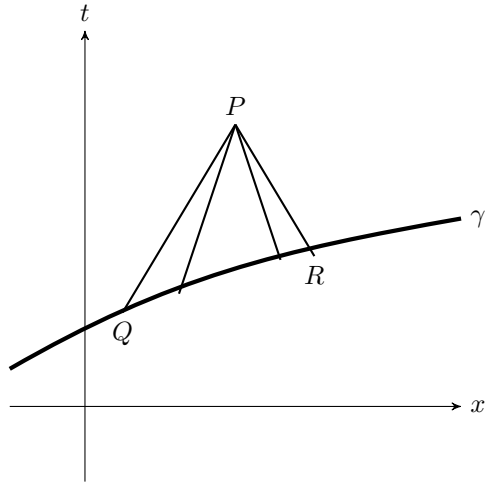

Let a $k$-th order totally hyperbolic linear PDE for a function $w$ of the two independent variables $x$ and $t$ be given.
Consider a smooth line $\gamma$ in the $x,t$ plane with the property of not being anywhere tangent to a characteristic direction. Assume, moreover, that on this line we have stipulated sufficiently regular values of the function $w$ and of its derivatives up to and including the order $k-1$ in a direction transversal to $\gamma$.\footnote{For the case of the system of second-order PDEs (\ref{beam-1}) and (\ref{beam0}) the data to be stipulated are the two functions $w$ and $\phi$ and their first derivatives in a direction transversal to $\gamma$.} The main result of the theory states that the solution at a point $P$ is completely and uniquely determined by the given data within the {\it domain of dependence} of $P$ obtained as the portion of $\gamma$ comprised between the two extreme intersections with the characteristic lines through $P$. For an equation with constant coefficients, the characteristics constitute $k$ independent systems of parallel straight lines. Figure \ref{fig:domain} shows the typical picture of the domain of dependence for the Timoshenko beam equation.

\begin{figure}
\begin{center}

\begin{tikzpicture}[scale = 0.5]
\draw[-stealth'] (-2,0) to  (10,0);
\draw[-stealth'] (0,-2) to  (0,10);
\draw[-] (6,0) to (6,10);
\node[right] at (10,0) {$x$};
\node[above] at (0,10) {$t$};
\draw[ultra thick] (0,0) to  (6,0);
\node[below] at (3,0) {$\gamma$};
\node[above] at (4,8) {$P$};
\draw[thick] (4,8) to (0,4);
\draw[thick] (4,8) to (6,6);
\draw[thick] (4,8) to (0,6);
\draw[thick] (4,8) to (6,7);

\end{tikzpicture}
\end{center}
\caption{Number of boundary conditions}
\label{fig:boundary}
\end{figure}
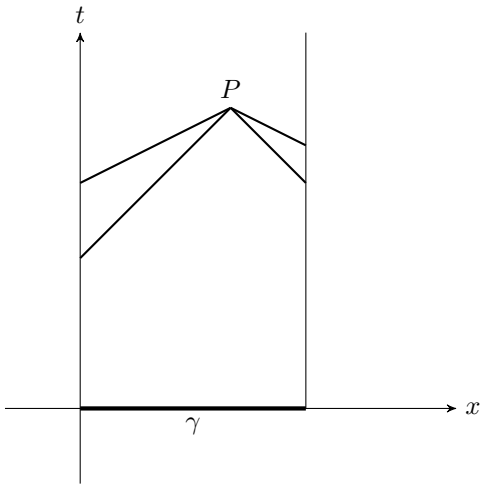

If the data and the solution are confined between two other non-characteristic lines, as shown in Figure \ref{fig:boundary} for the typical case of boundary conditions at the two ends of a finite beam, then additional conditions need to be specified for the function and its transverse derivatives on those lines. These conditions, however, cannot be prescribed arbitrarily. It can be shown\footnote{See John F. (1982), {\it Partial Differential Equations}, Springer, p 51.} that the number of conditions to be imposed on any such line must be equal to the number of intersections with this line of the characteristic lines issuing from $P$ toward $\gamma$. In the case of the Timoshenko beam, this implies that at each boundary exactly 2 boundary conditions must be specified. If 3 conditions were specified at one end and only 1 at the other end, this situation would in general imply an inconsistency of the initial data (overdetermined near one end and underdetermined near the other).

\section{Solution of the 2-2 problem}
At the free end ($x=L$) we have zero shear force and zero bending moment. The corresponding boundary conditions are
\begin{equation} \label{beam5}
w_x(L,t)-\phi(L,t)=0,
\end{equation}
and
\begin{equation} \label{beam6}
\phi_x(L,t)=0.
\end{equation}
Suppose now that the beam has been at rest for all times $t \le t_0$ and that for some time $t_1>t_0$ we want to have that
\begin{equation} \label{beam7}
w(L,t)=f(t-t_1),
\end{equation}
and
\begin{equation} \label{beam8}
\phi(L,t)=g(t-t_1),
\end{equation}
where $f(t)$ and $g(t)$ are sufficiently regular functions defined for all $t$ and vanishing identically for $t\le0$. The 2-2 problem consists of finding two inputs  (such as $w$ and $\phi$) at the held end so that the desired outputs ($f(t-t_1)$ and $g(t-t_1)$) are obtained at the free end while starting from vanishing initial conditions at all $t \le t_0$.

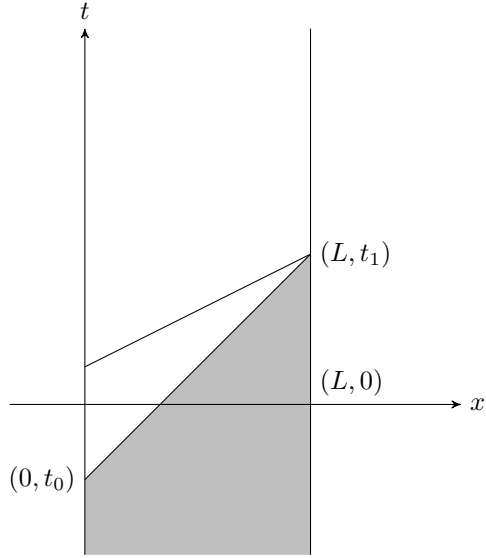
\begin{figure}
\begin{center}

\begin{tikzpicture}[scale = 0.5]
\path[fill=lightgray] (6,4) -- (0,-2) -- (0,-4) -- (6,-4) -- (6,4);
\draw[-stealth'] (-2,0) to  (10,0);
\draw[-stealth'] (0,-4) to  (0,10);
\draw[-] (6,-4) to (6,10);
\node[right] at (10,0) {$x$};
\node[above] at (0,10) {$t$};
\node[right] at (6,4) {$(L,t_1)$};
\node[left] at (0,-2) {$(0,t_0)$};
\node[above right] at (6,0) {$(L,0)$};
\draw[-] (6,4) to (0,-2);
\draw[-] (6,4) to (0,1);

\end{tikzpicture}
\end{center}
\caption{Solving the 2+2 problem}
\label{fig:twoplustwo}
\end{figure}

The proof that the solution to this problem exists and the actual procedure to construct it are based on the fact that the line $x=L$ is non-characteristic. Interchanging the roles of space and time,\footnote{Clearly, in more than one spatial dimension this procedure would be questionable, since the initial manifold has to be space-like.} the problem effectively becomes an `initial' value (Cauchy) problem. Indeed, since we have specified 4 independent data on this line, according to the fundamental theorem we can obtain a unique solution for the region $[0,L]\times \mathbb{R}$. This solution vanishes identically in the lower trapezoidal subregion shaded in Figure \ref{fig:twoplustwo}, namely the region below the `slow' characteristic line issuing from the point $(L,t_1)$. In particular, we obtain well-defined functions $F(t)=w(0,t)$ and $G(t)=\phi(0,t)$ which vanish for all $t<t_0=t_1- \sqrt{B}L$. The initial-boundary-value problem with the initial conditions
\begin{equation}
w(x,t_0)=\phi(x,t_0)=w_t(x,t_0)=\phi_t(x,t_0)=0,
\end{equation}
and with the boundary conditions (\ref{beam5}) and (\ref{beam6}) at $x=L$ and
\begin{equation}
w(0,t)=F(t),\;\;\;\;\;\;\;\;\;\;\phi(0,t)=G(t),
\end{equation}
is a well-posed problem. Its unique solution must perforce satisfy the desired displacement $f(t-t_1)$ and rotation $g(t-t_1)$ at $x=L$.

\section{Numerical example}

It is convenient to obtain a non-dimensional version of the equations of motion (\ref{beam-1}) and (\ref{beam0}). Defining the non-dimensional variables $\xi, \theta, \omega$ by
\begin{equation}
x=L \xi,\;\;\;\;\;\;t=L\sqrt{B} \tau,\;\;\;\;\;\;w=L \omega,
\end{equation} we obtain
\begin{equation}
\alpha \,\omega_{\tau\tau}=\omega_{\xi\xi}-\phi_\xi,
\end{equation}
and
\begin{equation}
\phi_{\tau\tau}=\phi_{\xi\xi}+\beta(\omega_\xi -\phi),
\end{equation}
where $\alpha=A/B$ and $\beta= L^2 C$ is a measure of the slenderness ratio of the beam. In most applications (that is, when the rotational inertia equals the mass density times the moment of inertia and when the shear stiffness is governed by a shear factor $k$) we obtain
\begin{equation}
\alpha=\frac{k}{2(1+\nu)}
\end{equation}
and
\begin{equation}
\beta= \alpha\,(L/\kappa)^2,
\end{equation}
where $\nu$ is Poisson's ratio and $\kappa$ is the radius of gyration of the cross section.

Interchanging the roles of space and time, we give as `initial' conditions on the line $\xi=1$ the following desired deflection
\begin{equation} \label{beam20}
\omega(1,\tau)=0.01 H(\tau + 5) (1-H(\tau-5)) \sin{\frac{2 \pi \tau}{5}},
\end{equation}
where $H(\cdot)$ is the Heaviside step function, a vanishing rotation (for lack of a better choice), namely,
\begin{equation}
\phi(1,\tau)=0,
\end{equation}
and zero shear and bending moment as stipulated by conditions (\ref{beam5}) and (\ref{beam6}). Notice that we have chosen our desired deflection at the free end ($\xi=1$) so that it starts to manifest itself, according to Equation (\ref{beam20}), at $\tau_1=-5$.

\begin{figure}[h]

\includegraphics[width=6.5in]{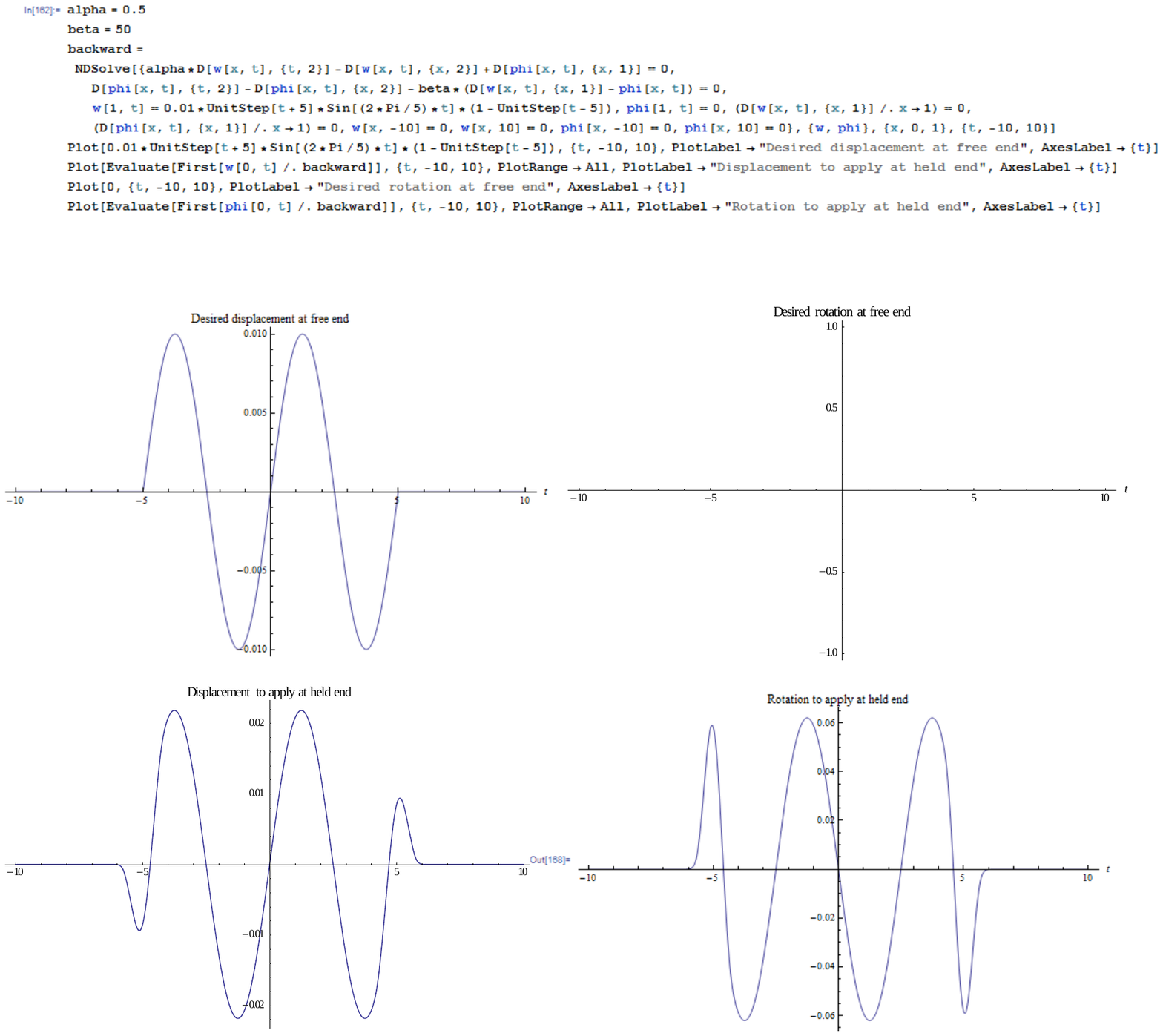}
\caption{Mathematica code and results}\label{fig:mathcode}
\end{figure}

To run this problem in a commercial software, we stipulate artificial vanishing `boundary' conditions of displacement and rotation at `time boundaries' located far enough from the domain of interest (at $t=\pm10$ in our example). For the values $\alpha=0.5, \beta=50$, the solution was obtained with Mathematica\textsuperscript{\textregistered}. The code is shown in Figure \ref{fig:mathcode}. Notice that the non-dimensional wave speeds are $\sqrt{2}$ and $1$, respectively, for the shear and the bending waves. Accordingly, the non-dimensional time for starting to impose the displacement and rotation on the held end ($\xi=0$), should be exactly equal to $\tau_0=\tau_1-1=-6$. And this is indeed the case in the numerical solution (which does not make use of characteristics), as can be observed in the graphs. The solution converges for a very wide range of values of the slenderness parameter $\beta$.

\section{The 1-1 problem}

On intuitive grounds it seems that, if the proverbial fishing rod is hinged to a support so as to permit only the variation of the angle at the held end, it should be possible to achieve any desired displacement at the free end provided one does not insist on imposing the rotation thereat. We call this the 1-1 problem. The intuitive reasoning is likely driven by a principle of conservation of the number of data necessary to solve a problem. It seems reasonable to conclude that all we are doing is exchanging the imposition of a desired rotation at the free end with a zero displacement at the other end.

In the 2-2 case, the solution was attainable because we managed to convert this problem into a 0+4 formulation, namely, when all 4 data were available along a single non-characteristic line ($\xi=1$). This is the so-called Cauchy problem, whose solution is guaranteed. To solve the 1-1 problem, on the other hand, we would have to resort to a 1+3 formulation. In other words, we would have to specify 3 data on the line $\xi=1$ (namely, the desired displacement and the vanishing of the bending moment and the shear force) and 1 on the line $\xi=0$ (namely, the vanishing of the displacement). Thus, it appears that, as we already pointed out in Section \ref{sec:wellposedness}, there will be a conflict between the initial conditions and the boundary conditions on either side of the rod. That this conflict may be resolvable should not come as a surprise, since we already encountered a similar conflict by imposing all 4 boundary conditions on a single end in the solution of the 2-2 problem. The key to the resolution resides in the fact that an arbitrary time delay is at our disposal. The details in the 1-1 case, however, are more subtle  and perhaps less convincing.

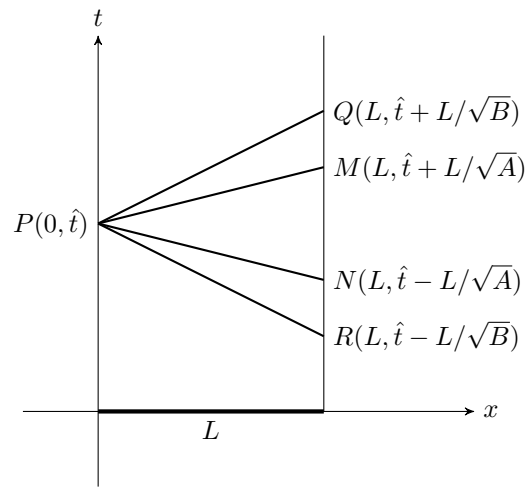
\begin{figure}
\begin{center}

\begin{tikzpicture}[scale = 0.5]
\draw[-stealth'] (-2,0) to  (10,0);
\draw[-stealth'] (0,-2) to  (0,10);
\draw[-] (6,0) to (6,10);
\node[right] at (10,0) {$x$};
\node[above] at (0,10) {$t$};
\draw[ultra thick] (0,0) to  (6,0);
\node[below] at (3,0) {$L$};

\draw[thick] (0,5) to (6,8);
\draw[thick] (0,5) to (6,2);
\draw[thick] (0,5) to (6,6.5);
\draw[thick] (0,5) to (6,3.5);
\node[left] at (0,5) {$P(0,{\hat t})$};
\node[right] at (6,8) {$Q(L,{\hat t}+L/\sqrt{B})$};
\node[right] at (6,2) {$R(L,{\hat t}-L/\sqrt{B})$};
\node[right] at (6,6.5) {$M(L,{\hat t}+L/\sqrt{A})$};
\node[right] at (6,3.5) {$N(L,{\hat t}-L/\sqrt{A})$};
\end{tikzpicture}
\end{center}
\caption{Argument for the 1-1 problem}
\label{fig:theoneoneproblem}
\end{figure}

Let a point $P$ be located on the line $x=0$, as shown in Figure \ref{fig:theoneoneproblem}. If data are to be specified on the line $x=L$, the domain of dependence thereat is dictated by the segment between the intersections, $Q$ and $R$, of this line with the two slower characteristics issuing form $P$.

The value of the solution at $P$, for the case of the homogeneous equation, depends exclusively on the data $w(L,t), w_t(K,t), \phi(L,t), \phi_x(L,t)$ for
${\hat t}-L/\sqrt{B} \le t \le {\hat t}+L/\sqrt{B}$. Of these data, we have specified the conditions of vanishing bending moment ($\phi_x(L,t) = 0$) and vanishing shear force ($w_x(L,t)=\phi(L,t)$). Moreover, we also specify the function $w(L,t)$ under the condition that it vanishes identically for all times prior to some value $t_1$, just as in the 2-2 problem. We have left the rotation $\phi(L,t)$ unspecified. The value of $w$ at point $P$, however, is known to vanish identically for all times (pinned end). Consequently, we have a functional equation for the missing function $\phi(L,t)$. Differentiating this equation with respect to $t$, we may obtain a more explicit expression in terms of the values of the functions $w(L,t)$ and $\phi(L,t)$ at the 4 points $M,N,Q,R$ only.

The argument just presented, which will be illustrated in the next section by means of an example, should be strengthened. In particular, the issue of continuous dependence of the solution (i.e., the rotation) on the boundary data (i.e., the displacement) deserves special attention.

\section{An example}

To obtain an explicit solution of the general 1-1 problem we would need a corresponding explicit expression of the solution of the Cauchy problem with `initial' data on the line $x=1$. This expression not being available for general values of the parameters $A$, $B$ and $D$ in Equation (\ref{beam1}), we will solve the problem only for the case $D=0$, namely, for the equation
\begin{equation} \label{beam30}
w_{xxxx}-(A+B)w_{xxtt}+AB w_{tttt}=0,
\end{equation}
whose physical meaning is of less practical interest, but whose characteristic polynomial is identical to that of the general case. From the point of view of the original pair of equations (\ref{beam-1}) and (\ref{beam0}), we have a weaker coupling resulting from neglecting the shear term in the second equation, that is,
 \begin{equation} \label{beam31}
 Aw_{tt}=w_{xx}-\phi_x,
 \end{equation}
 and
 \begin{equation}\label{beam32}
 B \phi_{tt}=\phi_{xx}.
 \end{equation}

The explicit solution of the Cauchy problem for these equations under the `initial' conditions stipulated by Equations (\ref{beam5}, \ref{beam6}, \ref{beam7}, \ref{beam8}) is given by
\begin{equation} \label{beam33}
w(y,t)=\frac{1}{2}(f(t+\sqrt{A}y)+f(t-\sqrt{A}y))- \frac{\sqrt{A}}{2(B-A)}\int\limits_{t-\sqrt{A}y}^{t+\sqrt{A}y}g(z)dz
+ \frac{\sqrt{B}}{2(B-A)}\int\limits_{t-\sqrt{B}y}^{t+\sqrt{B}y}g(z)dz,
\end{equation}
and
\begin{equation} \label{beam34}
\phi(y,t)=\frac{1}{2}(g(t+\sqrt{B}y)+g(t-\sqrt{B}y)),
\end{equation}
where we have shifted the origin to $x=L$ by introducing the variable
\begin{equation} \label{beam35}
y=x-L.
\end{equation}

We still need to determine the function $g$. To achieve this aim, we set the deflection to zero at $y=-L$, giving us the desired recursive functional equation. More explicitly, setting
\begin{equation} \label{beam36}
w_t(-L,t) = 0
\end{equation}
we obtain
\begin{eqnarray} \label{beam37}
g(t+\sqrt{B}L)&=& g(t-\sqrt{B}L)  \nonumber \\
&+& \sqrt{\frac{A}{B}}(g(t+\sqrt{A}L) - g(t-\sqrt{A}L)) \nonumber \\ &-& \frac{B-A}{\sqrt{B}}(f'(t+\sqrt{A}L)+f'(t-\sqrt{A}L)).
\end{eqnarray}
Having thus determined the missing `initial' condition, we have a legitimate Cauchy problem which can be solved for $w$ and $\phi$. The resulting function $w(-L,t)$ should clearly vanish, while the function $\phi(-L,t)$ provides the desired angle to be prescribed at the held end to produce the desired deflection at the free end. Notice that the fact that, by definition, the desired displacement $f$ vanishes identically for all $t \le 0$, guarantees the existence of a unique solution of Equation (\ref{beam37}), provided we impose the same condition on $g$. Similar considerations apply to the original Timoshenko equations, except for the fact that the recursion formula is not available explicitly.

\bigskip
\end{document}